\documentclass[
]{ceurart}

\usepackage{listings}
\begin{document}

\copyrightyear{2022}
\copyrightclause{Copyright for this paper by its authors.
  Use permitted under Creative Commons License Attribution 4.0
  International (CC BY 4.0).}

\conference{This paper was orally presented as one of the three out of nine accepted papers at the SIGIR 2023 Workshop on Reaching Efficiency in Neural Information Retrieval (ReNeuIR 2023), July 23-27, 2023, Taipei}
\title{Towards Consistency Filtering-Free Unsupervised Learning for Dense Retrieval}

\tnotemark[1]
\tnotetext[1]{We release codes at https://github.com/Haoxiang-WasedaU/Towards-Consistency-Filtering-Free-Unsupervised-Learning-for-Dense-Retrieval}

\author[1]{Haoxiang Shi}[%
email=hollis.shi@toki.waseda.jp,
]
\cormark[1]
\address[1]{Waseda University, 3-4-1 Ookubo, Shinjuku-ku, Tokyo, 169-8555, Japan}
\address[2]{Yahoo Japan Corporation,
  Kioi Tower 1-3 Kioicho, Chiyoda-ku, Tokyo, 102-8282, Japan}

\author[2]{Sumio Fujita}[%
email=sufujita@yahoo-corp.jp,
]

\author[1]{Tetsuya Sakai}[%
email=tetsuyasakai@acm.org,
]

\cortext[1]{Corresponding author.}

\begin{abstract}
Domain transfer is a prevalent challenge in modern neural Information Retrieval (IR).
To overcome this problem, previous research has utilized domain-specific manual annotations and synthetic data produced by consistency filtering to finetune a general ranker and produce a domain-specific ranker.
However, training such consistency filters are computationally expensive,  which significantly reduces the model efficiency.
In addition, consistency filtering often struggles to identify retrieval intentions and recognize query and corpus distributions in a target domain.
In this study, we evaluate a more efficient solution: replacing the consistency filter with either direct pseudo-labeling, pseudo-relevance feedback, or unsupervised keyword generation methods for achieving consistent filtering-free unsupervised dense retrieval. Our extensive experimental evaluations demonstrate that, on average, TextRank-based pseudo relevance feedback outperforms other methods. Furthermore, we analyzed the training and inference efficiency of the proposed paradigm. The results indicate that filtering-free unsupervised learning can continuously improve training and inference efficiency while maintaining retrieval performance. In some cases, it can even improve performance based on particular datasets.
\end{abstract}


\maketitle

\section{Introduction}
Ad-hoc text ranking, which involves creating ordered lists of candidate documents from a large static corpus in response to user queries, is a fundamental task in Information Retrieval (IR).
On the other hand, the Language models (LMs) have become the cornerstone of language tasks, because of their extensive wide pre-trained knowledge~\cite{wang2023chatgpt,10.1162/tacl_a_00520,wang-etal-2022-clidsum,wang2023towards,liang-etal-2023-summary,Wang2021KnowledgeES,shi2020siamese}.
The application of an LMs to ad hoc ranking has yielded significant performance improvements~\cite{nogueira2019passage}, which have driven a paradigm shift~\cite{lin2021pretrained} in IR.  In this paradigm, the previous assumption was that domain transfer could 
be generalized from one domain to all other domains.
However, cross-domain retrieval typically does not satisfy this assumption, limiting the performance.

To enhance domain adaptability, a two-stage approach has been widely adopted~\cite{guo2020multireqa}.
In the first stage, a general ranker is trained on open-domain question–answer (QA) data pairs. The solutions of the second stage include: (1) finetune the general ranker using manually
annotated query–document pairs of the target domain and
(2) generating synthetic data (i.e., query–document pairs with noise)~\cite{alberti2019synthetic} in the target domain using a consistency filter for fine-tuning. However, both methods are inefficient during the second stage.
Human labeling requires domain expertise and is time-consuming and expensive.
Additionally, the consistency filter usually relies on an external question–answer model trained by QA pairs. The additional training of the consistency filter is computationally expensive (for example, the T5 base model has 222.9M parameters)\cite{shi2022layerconnect}.

Furthermore, although the requirement of the labeled pairs is eliminated, the retrieval intents, query and corpus distributions are not always well recognized, given a target domain.~\citet{dai2022promptagator} then showed that using a generator trained by a few of labeled query-document pairs in the target domain can address these problems.

In this study, we aim at preserving the advantages of the two methods while circumventing their disadvantages. This involves abandoning both the external dependency and the annotation requirements, in addition to increasing the efficiency and meeting the real scenario in IR industry.
We aim to analyze whether a series of annotation-free and consistency filtering-free methods are efficient and effective for improving the domain transfer ability of the general ranker.

We use three mainstream extractive methods on domain-specific documents as a replacement for the filter in the second stage: (1) direct pseudo-labeling on the general ranker, (2) pseudo-relevance feedback~\cite{robertson2009probabilistic} (i.e., term frequency-inverse document frequency (TF-IDF)~\cite{sparck2004idf,salton1989automatic,lang1995newsweeder}, TextRank \cite{mihalcea2004textrank} and RAKE~\cite{rose2010automatic}) and (3) keywords generation using an unsupervised LM (i.e., KeyBERT~\cite{sharma2019self}). We then feed the extractive synthetic data as the positive samples into an unsupervised contrastive learning structure to finetune the general ranker~\cite{shi2020self}. The negative samples are from BM25 retrieval or random selection. 

We evaluate the performance of the extractive methods in the consistency filtering-free framework for dense retrieval on two specific-domain IR datasets: the WWW-4 web search dataset~\cite{sakai2020overview,mao2019overview} and the Robust04 News datasets~\cite{voorhees2003overview}. The performance is measured using four official evaluation metrics from the WWW-4 task~\cite{yamamoto2022overview}. The results suggest that utilizing TextRank-based pseudo relevance feedback achieved better performance compared to \emph{Direct Ranking} and \emph{Vanilla BM25} across all metrics on average, despite the lack of statistical significance.
Through an analysis of the training and inference efficiency of the proposed paradigm, we found that filtering-free unsupervised learning can continuously improve training and inference efficiency while maintaining retrieval performance. In fact, in some cases, it can even improve performance based on particular datasets.

We addressed the following research questions in this study:
\begin{itemize}
    \item Can cross-domain adaptation be realized using a consistency filtering-free unsupervised approach?
    \item How does the sampling configuration affect the performance of domain-specific ranking? 
    \item 
    How efficient is our method in terms of training and inference compared with PROMPTAGATOR~\cite{dai2022promptagator} and T5-based document expansion~\cite{nogueira2019document} by query prediction?

\end{itemize}

\section{Related Work}
\subsection{Dense Retrieval}
In contrast to discrete bag-of-word matching, dense retriever embeds queries and documents into a high-dimensional vector space and performs text matching in the embedding vector space, overcoming the vocabulary mismatch problem of sparse retrieval~\cite{li2022interpolate}.

Early dense retrievers utilize pre-trained LMs to encode queried and documents into bi-encoders to learn vector representations, via approximate nearest neighbor search \cite{karpukhin2020dense}. Although this training strategy is more efficient, the performance is limited because only calculating the inner product between the latent representation of the query and document as a loss~\cite{nogueira2019passage,karpukhin2020dense}. To address this limitation, \citet{khattab2020colbert} proposed an interaction framework that increases the number of interactions between queries and documents by enabling token-level interactions. Recent progress in the COIL framework by~\citet{gao-etal-2021-coil}, combines the exact matching and product of the latent presentations as the loss, thereby achieving both efficiency and effectiveness. 


\begin{figure}[t!]
\centering
\includegraphics[scale=0.7]{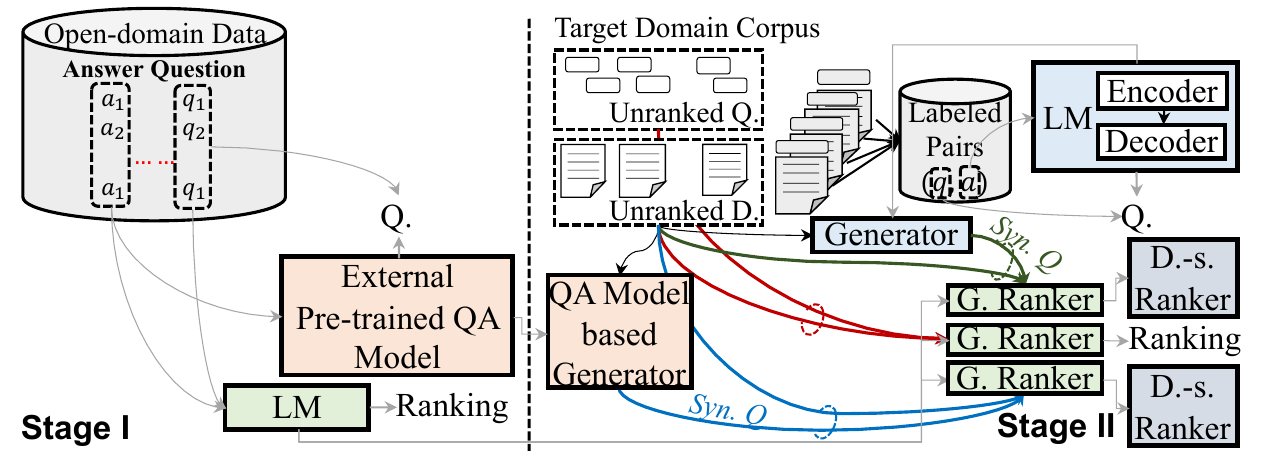}
\caption{Mainstream domain adaptation paradigms in Dense Retrieval.}
\label{methodcpr}
\end{figure}

\subsection{Domain Transfer via Consistency Filtering}
Domain adaptation remains a challenging topic in dense retrieval as current dense retrieval rankers are trained on open domain QA datasets, which provide limited insights into their out-of-distribution generalization capabilities. To enhance domain adaptability, using in-domain data to fine-tune a general ranker is the most effective method. However, owning to lack of domain-specific labeled data, various data augmentation methods were proposed, including query augmentation~\cite{ma2020zero} and document augmentation~\cite{raffel2020exploring}.

Among various query augmentation methods, the recently proposed technique of consistency filtering has been shown to improve the quality of generated queries by guaranteeing round-trip consistency; that is, the query must be answered by the passage it originated from \cite{alberti-etal-2019-synthetic}. In previous studies, to train the consistency filter, external QA models were preferred. For instance,~\citet{lewis2021paq} introduced probably asked questions to enhance the performance of the closed-book QA (CBQA) model. To remove the QA models, ~\citet{dai2022promptagator} trained an encoder-decoder as a consistency filter using a small group of in-domain query and document pairs. 

\section{Experiment}
Our task is to perform ad-hoc ranking on $D=\{q_i, d_j|i=1,2,\cdots,n; j=1,2,\cdots,m\}$, where $\{q_i\}$ and $\{d_j\}$ represent the queries and documents, respectively. To accomplish this, we first fine-tune a pre-trained language model (LM) using open domain question-answering (QA) data to obtain a general ranker $\textbf{R}$. We then further fine-tune the general ranker using a corpus specific to the target domain to obtain a domain-specific ranker $\tilde{\textbf{R}}$ based on $\textbf{R}$.

\subsection{Method}

\begin{figure}[t!]
\centering
\includegraphics[scale=0.8]{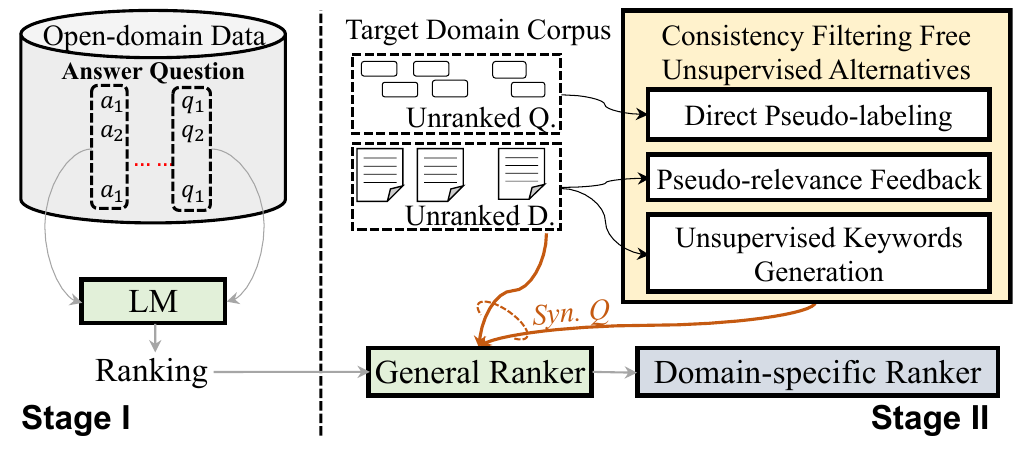}
\caption{The consistency filtering-free unsupervised dense retrieval paradigm.}
\label{method}
\end{figure}

In this method, we used the contextualized inverted list (COIL) framework~\cite{gao-etal-2021-coil} for both \textit{Stage I} and \textit{Stage II}, because COIL is the most efficient in an exact match. As shown in Fig.~\ref{method}, while keeping \textit{Stage I} unchanged, we modify the previous supervised cross-domain fine-tuning into an unsupervised \textit{Stage II}, and we adopt the widely-accepted contrastive learning loss in neural IR, which can be calculated as follows:

\begin{equation}
    \centering
    \mathcal{L} = -\log{\frac{\exp{[s(q,d^+)]}}{\exp{[s(q,d^+)]}+\sum^m_j{\exp{[s(q,d_{j}^{-})]}}}},
    \label{loss}
\end{equation}

\noindent where $s(\cdot,\cdot)$ denotes  the similarity which is the sum of the \texttt{[CLS]} product and the matching score. The \texttt{[CLS]} product is the product of the \texttt{[CLS]} representations of a query and a document output from the LM. The matching score is obtained by using each query token to look up its own inverted list and computing the vector similarity with document vectors stored in the inverted list, i.e., the contextualized exact lexical match. Previous methods distinguished a positive pairs and a negative pairs by using human notations or consistency filtering. By contrast, we analyze the performance of discriminating a positive pair $(q, d^+)$ in the following ways.

\textbf{Direct Pseudo-labeling.} We introduce a corpus $D^{*}=\{q_k, d_l\}$ that does not contain ranking information but is relevant to $D$. We feed each $q_k\in D^{*}$ into $\textbf{R}$ to retrieve the top-$U$ $d_l$ as positive pairs, which can be regarded as a pseudo-labeling of the retrieved documents by the corresponding query. This method can be understood as a domain-relevant data augmentation and is a query-triggered process. In an information retrieval (IR) system, it is easy to collect a wider-domain $D^{*}$ that includes the domain of $D$, a narrow-domain $D^{*}$ that belongs to the domain of $D$, or $D^{*}$ and $D$ can belong to the same wider domain.


\textbf{Pseudo-relevance Feedback.} Given a $d_l\in D^{*}$, we use the traditionally statistical methods to generate the top $V$ keywords $\{w_v\}$ of $d_l$ as the synthetic queries. Thus, $(\{w_v\}, d_l)$ is the positive pair. This approach is a document-triggered process and removes the requirements of $q_k\in D^*$. As for the selected methods to be analyzed, because TF-IDF is a basic method, we only expand the specific process of TextRank and RAKE. 

\begin{itemize}
\item \emph{TextRank}: The sentences in a document are parsed, and raw noun chunks and named entities are identified. The sentences are then iterated through to construct a lemma graph. Finally, \emph{PageRank}, a form of eigenvector centrality, is used to calculate ranks for each of the nodes, and the top-ranked phrases from the lemma graph are collected.

\item \emph{RAKE}: Keyword extraction begins with a document; the text is parsed into a set of candidate
keywords. After every candidate keyword is identified and the lemma graph of word co-occurrences is completed, a score is calculated for each candidate by summing the scores for candidate member words. After all candidate keywords are scored, the top $V$ scoring candidates are selected as keywords for the document.
\end{itemize}

\textbf{Unsupervised Keywords Generation.} This is also a document-triggered method. First, embeddings of a document $d_l\in D^{*}$ are extracted using BERT to obtain a document-level representation. Then, word embeddings are extracted for $N$-gram phrases. Finally, we use cosine similarity to find the words/phrases that are the most similar to the document $d_l$. The most similar words $\{w_v\}$ are then identified as the best description of the entire document.

Then, a negative sample $(q, d^-)$ is generated in the following ways. 
\begin{itemize}
    \item \emph{Random}: Given a pseudo-label or a synthetic $q_k$, we randomly select $m$ documents in $d_l\in D^*$.
    \item \emph{BM25}: For the direct pseudo-labeling, given a query $q_k\in D^*$, we use the BM25 method to retrieve the inverted rank of the documents $d_l\in D^*$, and select the top $m$ as the negative samples. Whereas, for the pseudo-relevance feedback and unsupervised keywords generation, a $d_l\in D^*$ is replaced by a synthetic query $\{w_v\}$, and the remaining process is kept the same.
\end{itemize}

The aforementioned methods use neither domain-specific labeled data nor
additionally trained collaborative filters. To verify the generalization ability, we evaluate these methods on different couples for $D$ and $D^*$.

\subsection{Dataset and Baseline}
We selected the following widely adopted large-scale ad hoc retrieval benchmarks from MSMARCO~\cite{craswell2021ms}, which contained three million English documents with an average length of approximately 900 tokens, as the open-domain QA pairs in \textit{Stage I}.
To evaluate \textit{Stage II}, We use two couples of $D$ and $D^*$, where the first couple is ($D$: WWW-4, $D^*$: WWW-2), and the second couple is ($D$: Robust$04$, $D^*$: TREC Washington Post Corpus). WWW-2 and WWW-4 are both the English web corpus~\cite{mao2019overview}. WWW2 is constructed from ClueWeb12-B13, whereas WWW-4 is constructed from Chuweb21 generated according to the April 2021 block of Common Crawl dataset5. The TREC Washington Post Corpus~\cite{bondarenko2018webis,soboroff2018trec} contains news articles and blog posts from January 2012 to December 2020. The Robust$04$ is generated from TREC disks 4 and 5, containing news data of five media including Financial Times Limited et al.
For WWW-2 and WWW-4 can be seen as the $D^*$ and $D$ are belong to the same wider domain, whereas for the TREC Washington Post Corpus and Robust$04$ can be seen $D^*$ is in a narrower domain than $D$.

Our baselines are as follows.
\begin{itemize}
    \item \emph{Direct Ranking}: Only the COIL-based \textit{Stage I} and no domain adaptation is conducted, and the general ranker retrieves the document ranking of a query $q_i\in D$. 
    \item \emph{Vanilla BM25}: Without the two-stage neural IR paradigm, the baseline retrieves the document ranking of a query $q_i\in D$.
\end{itemize}

\subsection{Setup}
\textbf{Main Experiment.} In \textit{Stage II}, each positive sampling method and each negative sampling methods are used in combination with each other ($10$ types in total). 
To enable a comparison with direct labeling, we followed the COIL's setting and fed one positive pair and $m=7$ negative pairs, which were randomly selected and retrieved using the BM25 method to retrieve irrelevant documents, into the general ranker.
In direct pseudo-labeling, we selected the top $U=5$ and $U=20$ results from the WWW-2 and Washington Post datasets, respectively, to serve as augmentation data for WWW-4 and Robust04.
The token and \texttt{[CLS]} dimensions are $n = 768$. A document and a query are truncated to 128 tokens and 16 tokens, respectively. For the training, the AdamW optimizer \cite{loshchilov2017decoupled} with a learning rate of $2e-6$, $0.1$ warm-up ratio, and linear learning rate decay.

\textbf{Oracle Experiments.} We further analyze the effect of different $U$ on the performance. As for pseudo-relevance feedback and unsupervised keywords generation, we further analyze the impact of different $V$ on the performance of TextRank.



\begin{table*}[t]
\centering
\resizebox{1.0\textwidth}{!}
{
\begin{tabular}{c|l|c|c|c|c}
\hline
& &\multicolumn{4}{c}{WWW-4} \\
\cline{3-6} ~
& ~ & \makecell[c]{nDCG@0010}  & \makecell[c]{Q@0010} & \makecell[c]{nERR@0010} & \makecell[c]{iRBU@0010} \\ 
\hline
\multirow{5}{*}{\rotatebox{90}{n: Random}} & p: Pseudo-labeling & 55.60 & 49.53 & 73.85 &  91.52 \\
 \cline{2-6} 
& p: TF-IDF      &  58.86 &  53.76 & 78.67  &   93.81   \\
& p: TextRank     &  \textbf{60.79} &  \textbf{56.65} & \textbf{79.81}  &   \textbf{94.00}    \\
& p: RAKE  &  56.08 &  51.57 & 73.08  &   92.78    \\
 \cline{2-6} 
& p: KeyBERT &        57.27    &  51.42 &  75.39 & 93.66     \\      
 \cline{1-6}
\multirow{5}{*}{\rotatebox{90}{n: BM25}} & p: Pseudo-labeling  & 58.71 & 54.62 & 75.92 & 91.09  \\
 \cline{2-6} 
& p: TF-IDF      &  57.05 &  51.97 & 76.19  &   93.38    \\
& p: TextRank     &  56.36 &  50.12 & 75.17  &   93.22     \\
& p: RAKE  &  58.22 &  52.66 & 76.72  &   94.34      \\
 \cline{2-6} 
& p: KeyBERT &  58.13 &  53.06 & 72.81  &   93.72    \\
\hline
\multicolumn{2}{c|}{Direct Ranking} &  57.50 &  53.97 & 72.09  &   92.13    \\
\multicolumn{2}{c|}{Vanilla BM25} &  51.70 &  48.06 & 67.11  &   89.20 \\
\hline
\end{tabular}
}
\caption{Experimental results on WWW-4, based on the COIL, where <p> and <n> refer to positive and negative sampling, respectively.}
\label{resultwww4}
\end{table*}

\begin{table*}[t]
\centering
\resizebox{1.0\textwidth}{!}
{
\begin{tabular}{c|l|c|c|c|c}
\hline
& & \multicolumn{4}{c}{Robust04} \\
\cline{3-6} ~
& ~ & \makecell[c]{nDCG@0010} & \makecell[c]{Q@0010} & \makecell[c]{nERR@0010} & \makecell[c]{iRBU@0010} \\ 
\hline
\multirow{5}{*}{\rotatebox{90}{n: Random}} & p: Pseudo-labeling & 42.18 & 30.73 & 56.55  & 67.02             \\
 \cline{2-6} 
& p: TF-IDF   &  41.81 &  30.02& 56.93  &   67.34           \\
& p: TextRank  & \textbf{43.07} &  31.39 & \textbf{58.08}  &  \textbf{68.69}         \\
& p: RAKE  &  42.01 &  30.89 & 55.98  &   67.05    \\
 \cline{2-6} 
& p: KeyBERT &  42.91   &  \textbf{31.46} &  57.97 & 68.18    \\      
 \cline{1-6}
\multirow{5}{*}{\rotatebox{90}{n: BM25}} & p: Pseudo-labeling   & 42.62 & 31.10 & 57.39 & 67.25             \\
 \cline{2-6} 
& p: TF-IDF      &  42.29 &  30.88 & 57.65  &   67.18            \\
& p: TextRank     &  42.63 &  31.03 & 57.38  &   68.07         \\
& p: RAKE  &  42.91 &  31.17 & 57.21  &   68.18   \\
 \cline{2-6} 
& p: KeyBERT &  42.63 &  31.03 & 57.38  &   68.07    \\
\hline
\multicolumn{2}{c|}{Direct Ranking} &  41.98 &  30.24 & 56.40  &   66.70    \\
\multicolumn{2}{c|}{Vanilla BM25} &  31.78 &  21.85 & 45.80  &   54.78    \\
\hline
\end{tabular}
}
\caption{Experimental results on Robust$04$, based on the COIL.}
\label{resultrobust}
\end{table*}

\section{Results}
\subsection{Can cross-domain adaption be realized via consistency
filtering-free unsupervised approach?
}

The results of the main experiment  experiments are presented in Tables~\ref{resultwww4} and ~\ref{resultrobust}. Each result is the average value of five runs. The following statements are based on average. The psuedo-labeling with \emph{BM25} negative sampling outperforms \textit{Direct Ranking} in most of the metrics on both $D$. By contrast, psuedo-labeling with \emph{Random} has a slight optimization effect on Robust$04$, however, basically shows negative optimization on WWW-4. TF-IDF with \emph{Random} and \emph{BM25} show a random performance and negative optimizations for both the datasets, respectively. RAKE with both the negative sampling methods slightly outperforms \textit{Direct Ranking} on Robust$04$ (except for nERR$@0010$), however, the opposite trend was observed for WWW-4, indicating a lack of generalization ability. TextRank with \emph{Random} outperforms other methods, and shows a stable performance for both the datasets, whereas the performance of TextRank with \emph{BM25} deteriorates to be close to \textit{Direct Ranking}. KeyBERT with \emph{BM25} outperforms that with \emph{Random} for WWW-4 , and the opposite results were obtained for Robust$04$. 
\\
\textbf{Statistical Significance Test.} We use randomized Tukey HSD test p-values~\cite{sakai2018laboratory}. The differences between the runs (excepting BM25) are not statistically significant.
The Tukey HSD test $p$-values are then shown in Table~\ref{hsdwww4} and Table~\ref{hsdrobost} for WWW-4 and Robust$04$, respectively. For only a system pair containing vanilla BM25, $p < 0.05$, whereas for other system pairs, all the $p > 0.05$, suggesting the improvements of our paradigms are not statistically significant.

These results indicate that it is feasible to remove the consistency filtering and domain-specific labeling and use unsupervised contrastive fine-tuning to achieve the domain transfer ability of a general ranker. Its performance depends on the combination of positive and negative sampling methods, suggesting that other unsupervised sampling methods are also worth exploring. In IR industry scenarios where there are many domains and many documents in each domain, this approach involves a trade-off between performance and efficiency compared to introducing complex and computationally expensive models. Possible reasons for the statistically insignificant improvements may be: (1) the query and corpus distributions are not significantly different between MSMARCO and our selected two groups of $D$ and $D^{*}$ ; (2) the selected $D$ and $D^{*}$ are not sufficiently strongly related in their corresponding domain; (3) the noise generated by the positive and negative sampling methods is relatively large, and there is still room for further optimization.

\begin{table*}[t]
\centering
\resizebox{1.0\textwidth}{!}
{
\begin{tabular}{c|c|c|c|c|c|c|c|c|c|c|c|c}
\hline
\makecell[c]{~} & \makecell[c]{D.R.} & \makecell[c]{P.-l.\& \\ RND} & \makecell[c]{P.-l.\& \\ BM25} & \makecell[c]{V. BM25} & \makecell[c]{TextRank\& \\ RND} & \makecell[c]{TextRank\& \\ BM25} & \makecell[c]{RAKE\& \\ RND} & \makecell[c]{RAKE\& \\ BM25} & \makecell[c]{KeyBERT\& \\ RND}	& \makecell[c]{KeyBERT\& \\ BM25}	& \makecell[c]{TF-IDF\& \\ RND}	& \makecell[c]{TF-IDF\& \\ BM25} \\
\hline
D.R. & ~ & 0.8379 & 0.4346 & 0 & 0.1808	& 0.6921 &	0.3484 & 0.5921 & 0.5605 & 0.4621 & 0.7588 & 0.4372 \\
P.-l.\&RND & 0.8379 & ~ & 0.5433 & 0 & 0.2328 & 0.8503 & 0.4434 & 0.7331 & 0.6963 & 0.5753 & 0.9145 & 0.5488 \\
P.-l.\&BM25 & 0.4346 & 0.5433 & ~ & 0 & 0.5692 &	0.6421 & 0.9002 & 0.7658 & 0.8045 & 0.9652 & 0.6155 & 0.98 \\
V. BM25	& 0	& 0	& 0	& ~ & 0	& 0 & 0	& 0 & 0 & 0 & 0 & 0 \\
TextRank\&RND & 0.1808 & 0.2328 & 0.5692 & 0 & ~ & 0.2737 & 0.6358	& 0.3644 & 0.393 & 0.5419 & 0.2772 & 0.5407 \\
TextRank\&BM25 & 0.6921 & 0.8503 & 0.6421 & 0 & 0.2737 & ~ & 0.5283 & 0.8653 & 0.8233 & 0.6785 & 0.9411 & 0.6508 \\
TF-IDF\&RMD & 0.3484 & 0.4434 & 0.9002 & 0 & 0.6358	& 0.5283	& ~ & 0.6537 & 0.6929 & 0.8647 & 0.512 & 0.512 \\
TF-IDF\&BM25	& 0.5921 & 0.7331 & 0.7658 & 0 & 0.3644 & 0.8653 & 0.6537	& ~ & 0.958	& 0.8027 & 0.8187 & 0.7791 \\
keyBERT\&RND & 0.5605 & 0.6963 & 0.8045 & 0 & 0.393 & 0.8233 & 0.6929 & 0.958	& ~	& 0.8415 & 0.7804 & 0.8191 \\
keyBERT\&BM25 & 0.4621 & 0.5753 & 0.9652 & 0 & 0.5419 & 0.6785 & 0.8647 & 0.8027 & 0.8415	& ~	& 0.649	& 0.9841 \\
RAKE\&RND & 0.7588 & 0.9145 & 0.6155 & 0 & 0.2772 & 0.9411 & 0.512 & 0.8187	& 0.7804 & 0.649 & ~ & 0.6234 \\
RAKE\&BM25 & 0.4372 & 0.5488 & 0.98 & 0 & 0.5407 & 0.6508 & 0.512 & 0.7791 & 0.8191 & 0.9841 & 0.6234 & ~ \\
\hline
\end{tabular}
}
\caption{Randomized Tukey HSD test p-values on WWW-4, where XXX\&BM25 refers to using XXX and BM25 for positive and negative sampling, respectively; XXX\&RND refers to using XXX and random for positive and negative sampling, respectively; D.R. is direct ranking, V. BM25 is vanilla BM25, and P.-l. is pseudo-labeling.}
\label{hsdwww4}
\end{table*}

\begin{table*}[t]
\centering
\resizebox{1.0\textwidth}{!}
{
\begin{tabular}{c|c|c|c|c|c|c|c|c|c|c|c|c}
\hline
\makecell[c]{~} & \makecell[c]{D.R.} & \makecell[c]{V. BM25} & \makecell[c]{P.-l.\& \\ RND} & \makecell[c]{P.-l.\& \\ BM25} & \makecell[c]{TextRank\& \\ RND} & \makecell[c]{TextRank\& \\ BM25} & \makecell[c]{RAKE\& \\ RND} & \makecell[c]{RAKE\& \\ BM25} & \makecell[c]{KeyBERT\& \\ RND}	& \makecell[c]{KeyBERT\& \\ BM25}	& \makecell[c]{TF-IDF\& \\ RND}	& \makecell[c]{TF-IDF\& \\ BM25} \\
\hline
D.R. & ~ &	0	&  0.9959 & 0.7651 & 0.7357	& 0.8643 & 0.9981 & 0.7823 & 0.7828 & 0.8643 & 0.8833 & 0.9691 \\
V. BM25 & 0 & ~ & 0 & 0 & 0 & 0 & 0 & 0 & 0 & 0 & 0 & 0 \\
P.-l.\&RND & 0.9959 & 0 & ~ & 0.8976 & 0.7337 & 0.8613 & 0.9978 & 0.78 & 0.7804 & 0.8613 & 0.8882 & 0.9652 \\
P.-l.\&BM25 & 0.7651 & 0 & 0.8976 & ~ & 0.653 & 0.9123 & 0.8543 & 0.7654 & 0.9345 & 0.8765 & 0.6754 & 0.8712 \\
TextRank\&RND & 0.7357 & 0 & 0.7337	& 0.653 & ~ &	0.8684 & 0.7367 & 0.9511 & 0.9524 & 0.8684 & 0.6289 & 0.7666 \\
TextRank\&BM25 & 0.8643 & 0 & 0.8613 & 0.9123 & 0.8684 & ~ & 0.8639 & 0.9167 & 0.9163 & 0.7515 & 0.8956 \\
RAKE\&RND & 0.9981 & 0 & 0.9978 & 0.8543 & 0.9511 &	0.9163 & ~ & 0.7829	& 0.7832 & 0.8639 & 0.8864 & 0.9676 \\
RAKE\&BM25 & 0.7823 & 0	& 0.78 & 0.7654 & 0.9511 & 	0.9167 & 0.7832	& ~ & 0.99 & 0.9167 & 0.673 & 0.8136 \\
KeyBERT\&RND & 0.7828 & 0 & 0.7804 & 0.9345 & 0.9524 & 0.9163 & 0.7832 & 0.99	& ~ & 0.9163 & 0.674 & 0.8138 \\
KeyBERT\&BM25 & 0.8643 & 0 & 0.8613	& 0.8765 & 0.8684 & 0.8732 & 0.8639 & 0.9167 & 0.9163 & ~ & 0.7515  & 0.8956 \\
TF-IDF &	0.8833 & 0 & 0.8882	& 0.6754 & 0.6289 & 0.7515 & 0.8864 & 0.673 & 0.674 & 0.7515 & ~ & 0.8956 \\
TF-IDF\&BM25 & 0.9691 & 0 & 0.9652 & 0.8712 & 0.7666	& 0.8956 & 0.9676	& 0.8136 & 0.8138 & 0.8956 & 0.8956 & ~ \\
\hline
\end{tabular}
}
\caption{Randomized Tukey HSD test p-values on Robust$04$.}
\label{hsdrobost}
\end{table*}

\begin{figure}[t]
\setlength{\abovecaptionskip}{0.1cm}
\centering
\includegraphics[scale=0.6]{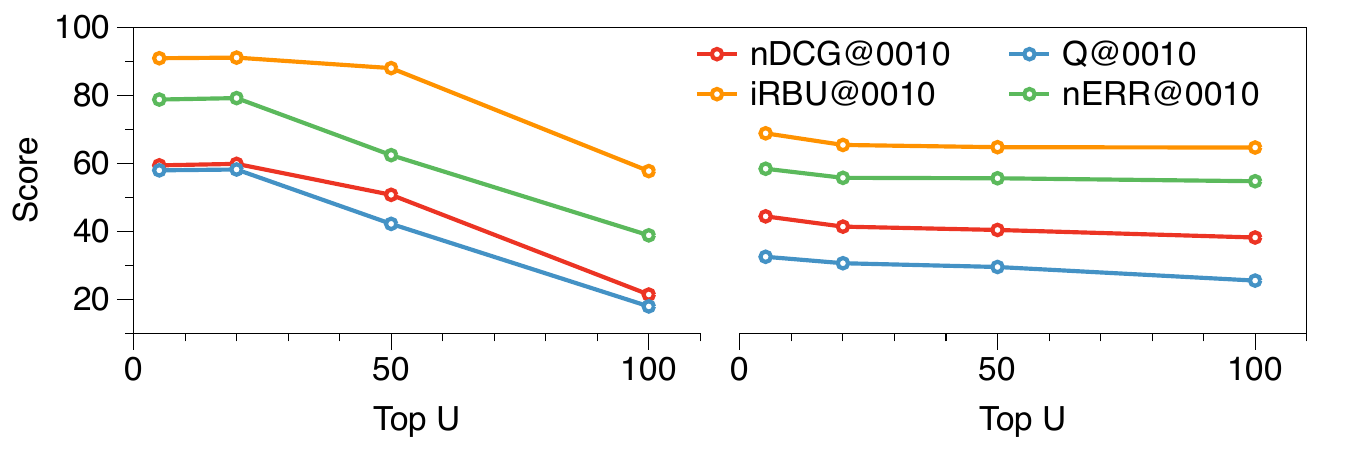}
\caption{The retrieval performance vs. different $U$ in pesudo-labeling using augment datasets WWW-2 and Washington Post on (a) WWW-4 and (b) Robust$04$.}
\label{U}
\end{figure}

\begin{figure}[t]
\centering
\setlength{\abovecaptionskip}{0.1cm}
\includegraphics[scale=0.6]{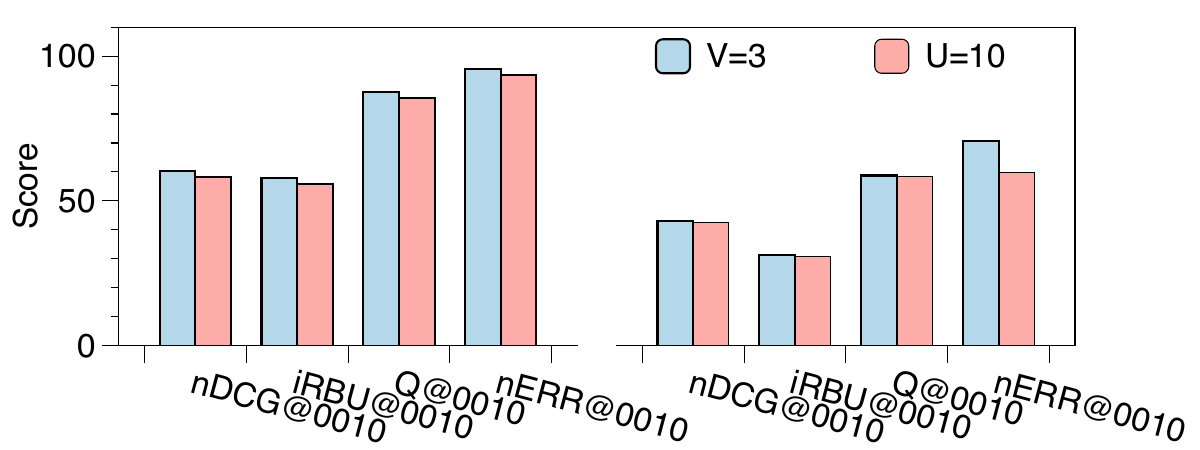}
\caption{The retrieval performance vs. different $V$ in TextRank on (a) WWW-4 and (b) Robust$04$.}
\label{V}
\end{figure}

\subsection{How sampling configuration impacts performance of 
domain-specific ranking
}

We answer this question from three perspectives.

\textbf{Impact on different of combinations of the sampling methods.} It can be observed that TextRank and pseudo-labeling exhibit opposite optimization effects under different negative sampling methods.This could be attributed to the fact that for pseudo-labeling, the query $q_i\in D$ does not have any noise, and hence the non-correlation of the negative samples is higher under the \emph{BM25} retrieval method compared to random sampling. This is because the latter method may introduce false negative samples.
In contrast, for TextRank, the extracted keywords serve as the synthetic query, which has noise and is more susceptible to false negative samples than random sampling. On the other hand, TextRank avoids the influence of open domain QA pairs on retrieval intents, query and document distribution in the target domain. We demonstrate that TextRank has better positive sampling ability than TF-IDF, RAKE, and KeyBERT. This leads us to conclude that minimizing the positive sampling noise and improving the non-correlation of the negative samples is a substantial factor for achieving successful domain adaptation performance.

\textbf{Impact on different $U$ in pseudo-labeling.} As shown in Fig.~\ref{U} (a) and (b), we find that in both $D$, the domain-adaptive performance deteriorates when $U$ is not the most suitable $U_s$. This $U$ cannot be too large, since too many documents of the same query increases the noise of the positive samples. Additionally, the performance on WWW-4 deteriorates more substantially than on Robust$04$ when $U > U_s$, suggesting that WWW-4 is more sensitive to this noise than Robust$04$. 
Therefore, it is important to determine the optimal value of $U$ for pseudo-labeling to achieve good results.

\textbf{Impact on different $V$ in TextRank.}: We take TextRank, which has the best performance, as a representative to analyze the impact of the keyword list length on \emph{Stage II}. As shown in Fig.~\ref{V} (a) and (b), the domain-adaptation performance deteriorates as $V$ is increased, on both $D$. This is because increasing the length of the may increase the noise of the synthetic query, thereby increasing the positive sampling noise. The results of adjusting both $U$ and $V$ simultaneously suggest the importance role of minimizing the positive sampling noise for ensuring good domain-adaptive performance.


\subsection{How our methods are efficient in terms of training and inference}

This study compares the training and inference efficiency of our consistency filtering-free unsupervised paradigm with PROMPTAGATOR~\cite{dai2022promptagator} and T5-based document expansion by query prediction~\cite{nogueira2019document}. Specifically, for model training, Compared to PROMPTAGATOR \cite{dai2022promptagator}, the proposed paradigm avoids the additional use of the T5 encoder (i.e., to be trained as the consistency filter), thereby achieving a reduction of 50$\%$ in total parameter count. Compare with to full T5 model-based document expansion by query prediction, the present paradigm results in a reduction of 67.8$\%$ in parameter count. Moreover, training a data generator with labeled data easily causes under-fitting, which suggests that the consistency filtering-free paradigm can solve such the problems.

With regard to model inference for query production (or keyword extraction in the proposed paradigm), TF-IDF-, TextRank- and RAKE-based consistency filtering-free unsupervised paradigms eliminate the need for an additional pre-trained LM, thereby resulting in faster inference speed compared with the both PROMPTAGATOR and full T5-based document expressions using the query predictions. Whereas, the pseudo-labeling- and KeyBERT-based paradigms rely on BERT model, thus the inference speed tends to be the same with the PROMPTAGATOR inference. However, the inference speed is till faster than the full T5 model-based inference. We also conduct an inference test on an Nvidia GTX3090 GPU card, and observe a $2.1\times$ acceleration in average when using TextRank-based paradigm, compared with the full T5 model-based paradigm.

\subsection{Failure analysis}
Based on the results of a static significance test, our proposed methods did not show significant improvements. In light of these findings, we conducted an in-depth analysis of several bad cases to investigate potential areas for improvement. Specifically, we evaluated the performance of our TextRank-based pseudo relevance feedback method and direct ranking by topics on the WWW4 dataset, using the nDCG@0010 scores of TextRank-based feedback. Based on these results, we identified the top $3$ queries that demonstrated superior and inferior nDCG@ $0010$ scores in TextRank-based feedback compared to Direct Labeling retrieval performance. Table \ref{badcase} presents our findings, which suggest that the TextRank-based method can outperform direct labeling for queries that are likely to retrieve clear answers.

\begin{figure}[t!]
\centering
\includegraphics[scale=0.6]{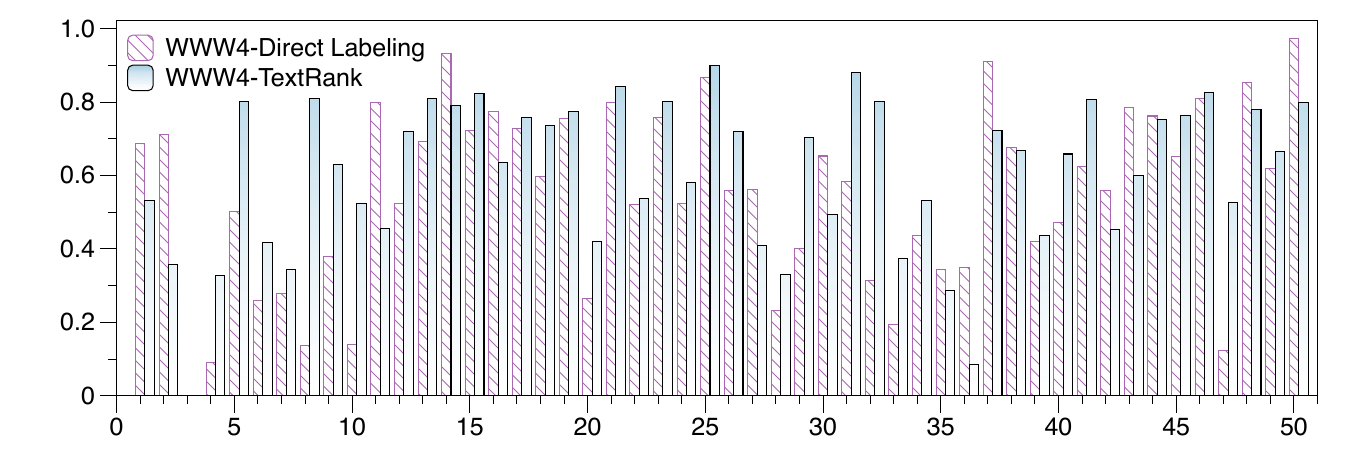}
\caption{The nDCG score by topics on the WWW4 dataset.}
\label{badcase}
\end{figure}

\begin{table*}[t]
\centering
\resizebox{0.6\textwidth}{!}
{
\begin{tabular}{ccc}
   \hline
   &QID & Content   \\
   \hline \hline
    Good retrieval topics
   & 0208 & cultural appropriation cases\\
   & 0210 & hypothermia treatment  \\
   & 0232 & chrome download windows  \\

   \hline
      Bad retrieval topics & 0202 & New Orleans restaurants  \\
   & 0211 & social anxiety  \\
   & 0236 & tennis score rules  \\
   \hline
  \end{tabular}
}
\caption{The top 3 queries demonstrate superior and inferior nDCG@0010
scores in TextRank-based feedback compared to Direct Labeling retrieval performance.}
\label{badcase}
\end{table*}

\section{Conclusion}
In this paper, we aim to eliminate the need for consistency filtering and manual labeling in the second stage of the previous two-stage methods, to achieve more efficient and low-cost domain adaptation in neural IR. We analyze three positive sampling methods combined with two negative sampling methods through unsupervised contrastive fine-tuning of the general ranker on the COIL framework.

The results of our study suggest that positive sampling using TextRank with \emph{Random} negative sampling outperforms the other methods on average, although these differences in performance are not statistically significant. These findings suggest that the consistency filtering-free unsupervised paradigm may have some ability for domain adaptation that relies on actual datasets.
Additionally, we find that during unsupervised fine-tuning of the domain-specific ranker in the second stage, it is a important factor to minimize the positive sampling noise and enhance the non-correlation of the negative samples in order to improve domain-adaptive ability. We additionally analyze the training and inference efficiency of the proposed paradigm. In training, the proposed paradigm saves 50\% and 67.8\% parameters, compared with PROMPTAGATOR and T5 model-based document expansion by query prediction, respectively. In inference, average $2.1\times$ acceleration has been tested compared to the full T5 model-based paradigm.
\section{Acknowledgment}
We sincerely thank Dr. Cen Wang, Mr. Jiaan Wang  and Mr. Qijiong Liu for their feedback
on our paper. Additionally, we would like to extend our thanks to the Sakai Lab from Waseda University for providing computational resources.

\bibliography{sample-1}


\end{document}